\documentclass[aps,prd,amsmath,amssymb,
nofootinbib,twocolumn,showpacs,preprintnumbers]{revtex4-1}
\pdfoutput=1

\newcommand\be{\begin{equation}}
\newcommand\ee{\end{equation}}
\newcommand\nono{\nonumber}
\newcommand\bse{\begin{subequations}}
\newcommand\ese{\end{subequations}}
\newcommand\bea{\begin{eqnarray}}
\newcommand\eea{\end{eqnarray}}

\usepackage[toc,page]{appendix}
\usepackage{graphics}      
\usepackage{graphicx}     
\graphicspath{{./GGDEdS2figs_pdfs_phiLOG_v2/}}
\usepackage{epsfig}
\usepackage{epstopdf}
\usepackage{caption} 
\usepackage{subcaption} 
\usepackage{float} 

\usepackage[table]{xcolor}    
\usepackage{url}           
\usepackage{bm}            
\usepackage{mathrsfs}

\usepackage{hyperref}
\usepackage{color}
\usepackage{soul}
\usepackage{tikz}
\usetikzlibrary{calc}
\usetikzlibrary{decorations.pathreplacing}

\usepackage{tikz-3dplot}

\numberwithin{equation}{section}

\newcommand\ringring[1]{
  {
   \mathop{\kern0pt #1}\limits^{
     \vbox to-1.85ex{
       \kern-2ex 
       \hbox to 0pt{\hss\normalfont\kern.1em \r{}\kern-.45em \r{}\hss}
       \vss 
     }
   }
  }
}

\DeclareMathOperator{\arcsinh}{arcsinh}

\begin{document}
\title{Generalized geodesic deviation in de Sitter spacetime}
\author{Isaac Raj Waldstein}
\email{isaac.waldstein@duke.edu}
\affiliation{Department of Physics and Astronomy, University of North Carolina at Chapel Hill, Phillips Hall CB 3255, Chapel Hill, North Carolina 27599, USA}
\author{J. David Brown}
\affiliation{Department of Physics, North Carolina State University, Raleigh, NC 27695}
\date{\today}

\begin{abstract}
The geodesic deviation equation (GDE) describes the tendency of objects to accelerate towards or away from each other due to spacetime curvature. 
The GDE assumes that nearby geodesics have a small rate of separation, which is formally treated as the same order in smallness as the separation itself. 
This assumption is discussed in various papers but is not articulated in any standard textbooks on general relativity.
Relaxing this assumption leads to the generalized geodesic deviation equation (GGDE).
We elucidate the distinction between the GDE and the GGDE by explicitly computing the relative acceleration between timelike geodesics in two--dimensional de Sitter spacetime.
We do this by considering a fiducial geodesic and a secondary geodesic (both timelike) that cross with nonzero speed. 
These geodesics are spanned by a spacelike geodesic, whose tangent evaluated at the fiducial geodesic defines the separation. 
The second derivative of the separation describes the relative acceleration between the fiducial and secondary geodesics. 
Near the crossing point, where the separation between the timelike geodesics is small but their rates of separation can be large, we show that the GGDE holds but the GDE fails to apply. 
\end{abstract}

\maketitle
\makeatletter
\let\toc@pre\relax
\let\toc@post\relax
\makeatother

\section{Introduction}\label{sec1}

In general relativity, the geodesic deviation equation\footnote{Also known as the Jacobi equation.} (GDE) determines the relative acceleration between neighboring, freely falling objects.
Synge and Schild \cite{Synge1952, Synge1960} introduced the GDE into the body of fundamental concepts in general relativity. 
The GDE quantifies the evolution of the separation between nearby geodesics.
 
As detailed in Ref.~\cite{Vines2015}, there are many different ways to define the separation between geodesics. 
This is because there is no unique way of defining a correspondence between points on nearby geodesics. 
In  many textbooks on general relativity \cite{Weinberg1972, MTW1973, Peebles1993, OR1994, Hartle2003, Hobson2006, Schutz2009}, the correspondence is fixed by choosing points with common values of affine parameter $\tau$. 
The difference between coordinate values of these points  defines the separation vector $S^\kappa(\tau)$. 

Other textbooks \cite{Synge1952, Synge1960, HawkingEllis1975, Carroll2004, Wald1984, Pad2010} consider a continuous family of geodesics, with each member of the family parametrized by $\tau$. In this case, the separation vector is defined as the tangent to the curves $\tau = \mathrm{constant}$.

In each of these cases, we identify the separation vector $S^\kappa(\tau)$ as a tangent vector  at a point along a ``fiducial" geodesic, defining the  separation between the fiducial geodesic and a nearby secondary geodesic. 
The resulting GDE is
\be\label{GDE}
\frac{D^2 S^\kappa}{d\tau^2} = -  R^\kappa_{\lambda \mu \nu}\,T^\lambda\,S^\mu\,T^\nu +  \mathcal{O}\left(S, \frac{DS}{d\tau}\right)^2  ,
\ee
where $\mathcal{O}\left(S, {DS}/{d\tau}\right)^2$ denotes error terms with two or more factors of $S^\kappa$ and/or ${DS^\kappa}/{d\tau}$. 
Here, $T^\nu$ is the tangent vector to the fiducial geodesic and $D/d\tau$ denotes the 
covariant derivative along the fiducial geodesic. 
Thus, $DS^\kappa/d\tau$ is the rate of separation (or relative velocity)~\cite{Vines2015} between the fiducial and secondary geodesic.

The GDE~\eqref{GDE} says that the relative acceleration between geodesics is proportional to the Riemann tensor $R^\kappa_{\lambda \mu \nu}$. 
Without the error terms, Eq.~\eqref{GDE} is the form of the geodesic deviation equation most commonly found in standard textbooks on general relativity \cite{MTW1973, Hartle2003, Hobson2006, Carroll2004, Wald1984, Pad2010}. 
The GDE treats the separation $S^\kappa$ as small compared to $\mathcal{R}$ (the radius of curvature of spacetime). 

Standard textbooks do not expound that the GDE treats the rate of separation $DS^\kappa/d\tau$ as the same order in smallness compared to $c$ (the speed of light in vacuum) as the separation itself.\footnote{See for example Refs.~\cite{Synge1952, Synge1960, Weinberg1972, MTW1973, HawkingEllis1975, Wald1984, Peebles1993, Hartle2003, Carroll2004, Hobson2006, Schutz2009, Pad2010}. Perhaps one exception is Ohanian and Ruffini \cite{OR1994} who state that the separation and rate of separation are assumed small, but provide no discussion or explanation.}  Nor do they explain that we can relax the assumption that the neighboring geodesics have small rates of separation compared to $c$. That is, we keep terms through order one in $S^\kappa$ and we keep all orders in ${DS^\kappa}/{d\tau}$.  This yields the generalized geodesic deviation equation (GGDE). 
The GGDE was introduced by Hodkingson \cite{Hod1972} and studied by Mashhoon and Chicone \mbox{\cite{Mashhoon1975,Mashhoon1977, Chicone2002, Chicone2006, Chicone2006v2}} and other authors \cite{Li1979, Ciufolini1986, Ciufolini1986v2, Perlick2008}.

The \emph{isochronous correspondence} requires points on two neighboring, affinely parametrized geodesics to be linked by a connecting geodesic at the same value of affine parameter.
In the isochronous correspondence, the GGDE takes the form 
\bea\label{ggdedS2v1}
&& \frac{D^2S^\kappa}{d\tau^2}  =  -R^\kappa_{\lambda \mu \nu}\,T^\lambda\,S^\mu\,T^\nu\\\nono  
&&- R^\kappa_{\lambda \mu \nu}\,\left[ 2\,\frac{DS^\lambda}{d\tau}\, S^\mu\, T^\nu +  \frac{2}{3}\, \frac{D S^\lambda}{d\tau}\, S^\mu\,\frac{DS^\nu}{d\tau}\right]  +  \mathcal{O}(S)^2 .
\eea
Like the geodesic deviation equation \eqref{GDE}, the generalized geodesic deviation equation \eqref{ggdedS2v1} is linear in the separation $S^\kappa$ but it is nonlinear in relative velocity $DS^\kappa/d\tau$. 
The GGDE generalizes the GDE to higher orders by accounting for arbitrary relative velocities between neighboring geodesics \cite{Vines2015,Hod1972}.

Chicone and Mashhoon \cite{Chicone2002, Chicone2006, Chicone2006v2} used the GGDE to analyze fleets of test particles in Kerr spacetime. 
They studied applications to astrophysical jets, and to tidal dynamics in metrics with high degrees of symmetry.
Other applications of the GGDE \cite{Li1979,Ciufolini1986, Ciufolini1986v2,Perlick2008} include a scheme proposed by Ciufolini and Demianski for measuring spacetime curvature. 

As opposed to the isochronous correspondence, Chicone and Mashhoon \cite{Chicone2006} investigated solutions to the GGDE in the normal correspondence.\footnote{With the normal correspondence, the separation vector is fixed to be everywhere orthogonal to the fiducial geodesic.} 
Specializing to one-dimensional motion along an axis of symmetry, they used spherical Fermi coordinates to exactly compute the radial motion of test particles in de Sitter spacetime. 
They wrote the metric in terms of the Hubble rate for cosmological expansion and found that the first-order tidal terms in the GGDE approximate the exact radial motion of the test particles.

More recently, Vines used the exponential map to derive a manifestly covariant form of the geodesic deviation equation in the isochronous correspondence that is valid to all orders in the separation and relative velocity \cite{Vines2015}. 
This result reduces to Eq.~\eqref{ggdedS2v1} at first order in the separation.
Building on Ref.~\cite{Vines2015}, Flanagan et.~al \cite{Flanagan2019} used Jacobi propagators to construct
persistent gravitational--wave observables. 
Their work generalizes the geodesic deviation equation to allow for acceleration, which can arise due to spin or self--force effects.
Meanwhile, Puetzfeld and Obukhov derived an exact form of the geodesic deviation equation that extends previous generalizations and qualitatively agrees with the results of Vines in certain cases \cite{Puetzfeld2016}. 
By applying the results in~\cite{Puetzfeld2016}, Obukhov and Puetzfeld offer 
new prescriptions for measuring the gravitational field in Ref.~\cite{Obukhov2019}.

In this paper, we focus on the GGDE in the isochronous correspondence. 
This choice of parametrization leads to the simplest forms for the GDE and GGDE \cite{Vines2015}. 
We demonstrate the distinction between the GDE and the GGDE by explicitly computing the relative acceleration $D^2 S^\kappa/d\tau^2$ between timelike geodesics in two--dimensional de Sitter spacetime. 
When the separation between geodesics is small but their rate of separation is large, we show that the relative acceleration is given correctly by the GGDE~\eqref{ggdedS2v1}. 
We substantiate this result by directly computing the effect of the error terms in the GDE using an orthonormal basis.
The textbook geodesic deviation equation \eqref{GDE} only applies when the relative velocity is small. 

This paper elucidates the difference between the GDE and the GGDE in the context of a tractable example.
In Sec.~\ref{sec2}, we briefly outline the derivation of the GGDE given by Mullari and Tammelo in Ref.~\cite{Mullari2000}. 
Section~\ref{sec3} introduces the de Sitter geometry in which we compute the relative acceleration between geodesics.
In Sec.~\ref{sec4}, we explicitly solve for the spacelike geodesics that connect the fiducial and secondary timelike geodesics in the isochronous correspondence. 
In Sec.~\ref{sec5}, we directly compute the relative acceleration between two geodesics that are close to passing one another at high speed. 
We explicitly verify that the GGDE yields the correct relative acceleration, whereas the GDE does not. 
Thus, the relative velocity terms that are present in the GGDE but ignored in the GDE are non-negligible. 
The sign conventions of Misner, Thorne, and Wheeler \cite{MTW1973} are used throughout this paper.

\section{The GGDE}\label{sec2}

In both the generalized geodesic deviation equation  (\ref{ggdedS2v1}) and the geodesic deviation equation (\ref{GDE}), the second covariant derivative of the separation vector (or relative acceleration) $A^\kappa  \equiv {D^2S^\kappa}/{d\tau^2}$
is assumed to be of the same order in smallness as $S^\kappa$ itself. 
All terms in Eqs.~(\ref{ggdedS2v1}) and~(\ref{GDE}) are to be evaluated on the fiducial geodesic. 
If we treat the relative velocity terms as the same order in smallness as the separation itself, then the GGDE in (\ref{ggdedS2v1}) reduces to the GDE in (\ref{GDE}), as required. 

By today's standards, the original derivation of the GGDE by Hodgkinson \cite{Hod1972} is mathematically cumbersome. 
Mullari and Tammelo give a shorter and cleaner derivation of the GGDE in Ref.~\cite{Mullari2000}.\footnote{To our knowledge, Ref.~\cite{Mullari2000} has been overlooked in the literature on geodesic deviation.}
We outline their derivation here, using our notation.

\begin{figure}[h]
\centering
\begin{tikzpicture}
\begin{scope} [ ]
\draw [->][] plot [smooth] coordinates { (0.3,-0.3) (0.65,0.25) (0.35,1.25)  (0.2,2.0) };
\node at (-0.2,-0.55) {$G$: $x^\kappa = \sigma^\kappa(\tau)$};
\node at (0.32,0.20) {$\tau$};
\node (dot) at (0.36,1.2) {};
\draw [fill] (dot) circle (0.04);
\node at (0.14,1.2) {$P$};
\node at (-0.77,1.2) {$\xi^\kappa(s=0):$};
\draw [->][] (0.36,1.22)--(1.0,1.3) node [] {};
\node [] at (0.58, 1.42) {$S^\kappa$};
\node at (1.1,0.94) {$s$};
\end{scope}
\draw [] [] plot [smooth, tension=0.6] coordinates {(dot)  (1.0, 1.2) (1.5,0.7)};
\begin{scope}[ shift={(1.0,0)} ]
\draw [->] [] plot [smooth] coordinates { (0.5,-0.5) (0.65,0.25) (0.35,1.25)  (0.5,2.2) };
\node at (1.2,-0.8) {$\bar{G}$: $x^\kappa = \bar{\sigma}^\kappa(\bar{\tau})$};
\node at (0.80,0.05) {$\bar{\tau}$};
\node (dot) at (0.51,0.7) {};
\draw [fill] (dot) circle (0.04);
\node at (0.70,0.9) {$\bar{P}$};
\node at (1.63,0.9) {$ :\xi^\kappa(s=1)$};
\end{scope}
\end{tikzpicture}
\caption{Two closely separated geodesics $G$ (left) and $\bar{G}$ (right) linked by a ``connecting" geodesic (center). All geodesics shown in Fig.~\ref{fig: geodesicsv1} are affinely parametrized.} 
\label{fig: geodesicsv1}
\end{figure}
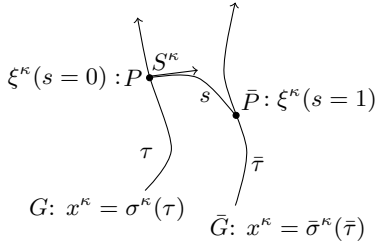

Consider two nearby geodesics $G$ (fiducial) and $\bar{G}$ (secondary)
with affine parameters $\tau$ and $\bar{\tau}$  and coordinates $x^\kappa = \sigma^\kappa(\tau)$ and $x^\kappa = \bar{\sigma}^\kappa(\bar{\tau})$, respectively. 
See Fig.~\ref{fig: geodesicsv1}.
We assume that there exists a mapping $\bar\tau = \bar\tau(\tau)$
from points $P$ on $G$ to corresponding points $\bar P$ on $\bar{G}$.
Each pair of corresponding points $P$ and $\bar P$ is connected by a geodesic 
\be\label{geodesiclinesL}
\mathcal{C}: x^\kappa = \xi^\kappa(s),
\ee
with coordinates $\xi^\kappa(s)$.
Here, $s$ is an affine parameter that ranges from $0$ to $1$ so that
\be\label{PbarPrelations}
P: \xi^\kappa(s=0) = \sigma^\kappa(\tau), \quad \bar{P}: \xi^\kappa(s=1) = \bar{\sigma}^\kappa(\bar{\tau}(\tau)).
\ee 
We now expand the coordinates of $\bar{P}$ about $s=0$ and use the relations \eqref{PbarPrelations} to obtain
\bea\label{expansionv1}
\bar\sigma^\kappa(\bar\tau(\tau)) & = & \sigma^\kappa(\tau) + \frac{d\xi^\kappa(s)}{ds}\Bigg|_{s=0}\, \\\nono
&& + \,\,\frac{1}{2}\,\frac{d^2\xi^\kappa(s)}{ds^2}\Bigg|_{s=0} + \frac{1}{6}\frac{d^3\xi^\kappa(s)}{ds^3}\Bigg|_{s=0}   + \cdots .
\eea
The derivative ${d\xi^\kappa(s)}/{ds}$ evaluated at $s = 0$ defines the separation vector
\be\label{Skappa}
S^\kappa(\tau) \equiv \frac{d\xi^\kappa(s)}{ds}\Bigg|_{s=0}, 
\ee
between the fiducial and secondary geodesics. 
As shown in Fig.~\ref{fig: geodesicsv1}, $S^\kappa(\tau)$ is tangent to the ``connecting" geodesic with coordinates $\xi^\kappa(s)$. 
The separation vector $S^\kappa(\tau)$ is a function of $\tau$ because it depends on the chosen point $P$ on the fiducial geodesic. 

Using the fact that $\xi^\kappa(s)$ is a geodesic, 
\be\label{connectinggeodesic}
\frac{d^2\xi^\kappa(s)}{ds^2} = - \Gamma^\kappa_{\alpha \beta}(\xi(s))\,\frac{d\xi^\alpha}{ds}\, \frac{d\xi^\beta}{ds},
\ee
we can rewrite the series in (\ref{expansionv1}) as 
\bea\label{expansionv2}
\bar{\sigma}^\kappa(\bar{\tau}(\tau))  & = &  \sigma^\kappa(\tau) + S^\kappa - \frac{1}{2}\,\Gamma^\kappa_{\alpha \beta}(\sigma(\tau))\, S^\alpha \,S^\beta \\\nono
&& -\frac{1}{6}\,\partial_\nu\,\left\{\Gamma^\kappa_{\lambda \mu}(\sigma(\tau))\right\}\, S^\lambda \, S^\mu S^\nu\\\nono
&& +\frac{1}{3}\, \Gamma^\kappa_{\pi \nu}(\sigma(\tau)) \,\,\Gamma^\pi_{\lambda \mu}(\sigma(\tau))\,\, S^\lambda\, S^\mu \, S^\nu + \mathcal{O}(S)^4,
\eea
keeping terms through third order in the separation vector.

Next, following Mullari and Tammelo, we differentiate Eq.~(\ref{expansionv2}) twice with respect to $\tau$ and use
\be\label{chainrule}
\frac{d}{d \bar{\tau}} = \frac{d \tau}{d\bar{\tau}}\, \frac{d}{d \tau}. 
\ee
During this process, we use the fact that $\sigma^\kappa(\tau)$ and $\bar{\sigma}^\kappa(\bar{\tau})$ are geodesics
\bse
\bea
\frac{d^2\sigma^\kappa}{d\tau^2} &=& - \Gamma^\kappa_{\alpha \beta}(\sigma(\tau))\,\frac{d\sigma^\alpha}{d\tau}\, \frac{d\sigma^\beta}{d\tau},\\
\frac{d^2\bar{\sigma}^\kappa}{d\bar{\tau}^2} &=& - \Gamma^\kappa_{\alpha \beta}(\bar{\sigma}(\bar{\tau}))\,\frac{d\bar{\sigma}^\alpha}{d\bar{\tau}}\, \frac{d\bar{\sigma}^\beta}{d\bar{\tau}},
\eea
\ese
and expand the Christoffel symbols at $\bar P$ as 
\be\label{expansionv3}
\Gamma^\kappa_{\alpha \beta}(\bar{\sigma}(\bar{\tau})) = \Gamma^\kappa_{\alpha \beta}(\sigma(\tau)) + \partial_\nu \left\{\Gamma^\kappa_{\alpha \beta} \right\}\Big|_{x = \sigma(\tau)}\, S^\nu(\lambda) + \mathcal{O}(S)^2.
\ee
We also rewrite terms involving $d\bar\sigma^\alpha/d\bar\tau$ using the first 
derivative of Eq.~\eqref{expansionv2}. 
In the resulting expression, ordinary derivatives are traded for covariant derivatives along the fiducial curve. We assume that ${D^2 S^\kappa}/{d\tau^2}$ is the same order in smallness as $S^\kappa$ itself, but that ${DS^\kappa}/{d\tau}$ is finite. 
If we restrict to the isochronous correspondence where $\tau = \bar{\tau}$, then $d\tau/d\bar\tau$ is unity and $d^2\tau/d\bar\tau^2$ vanishes.  
This leads to the result of Mullari and Tammelo \footnote{Equation~\eqref{MT6} is identical to Eq.~(6) of Mullari and Tammelo~\cite{Mullari2000}, specialized to the isochronous correspondence. 
In Ref.~\cite{Mullari2000}, Eq.~(6) should follow from Eqs.~(4) and (5). However, Eq.~(4) is missing a term  $-2\, ({dS^\lambda}/{d\tau}) S^\mu\, \Gamma^\kappa_{\alpha \beta}\, \Gamma^\beta_{\lambda \mu}\, T^\alpha$ (in our notation) and Eq.~(5) should have a plus sign, not a minus sign, in front of the term $\Gamma^\kappa_{\lambda \mu} ({D^2S^\lambda}/{d\tau^2}) S^\mu$.}
\bea\label{MT6}
&&\frac{D^2 S^\kappa}{d\tau^2} = -R^\kappa_{\lambda \mu \nu}\, T^\lambda\, S^\mu\, T^\nu \\\nono
&& - \,\,  R^\kappa_{\lambda \mu \nu} \left[2\,\frac{D S^\lambda}{d\tau}\, S^\mu\, T^\nu
+  \frac{2}{3}\, \frac{D S^\lambda}{d\tau}\, S^\mu\,\frac{DS^\nu}{d\tau} \right] + \,\,  \mathcal{O}(S)^2,
\eea
and matches the GGDE in (\ref{ggdedS2v1}).

\section{de Sitter Spacetime}\label{sec3}

To study geodesic deviation in curved spacetime, we will work in de Sitter spacetime, the maximally symmetric spacetime with positive curvature.
The metric is defined by embedding a hyperboloid
\be\label{hyperboloid}
R^2  =  -T^2 + X^2 + Y^2,
\ee
in a $(2+1)$--dimensional Minkowski spacetime with line element $ds^2 = -dT^2 + dX^2 + dY^2$. 
With coordinates $\left\{t, \phi\right\}$ satisfying 
\bse\label{deSitterv2}
\bea
T & = & R\, \sinh\left(\frac{t}{R}\right)\, \cosh(\phi),\\
X & = & R\, \cosh\left(\frac{t}{R}\right),\\
Y & = & R\, \sinh\left(\frac{t}{R}\right)\sinh(\phi),
\eea
\ese
the metric is given by 
\be\label{metricdS2}
ds^2  =  -dt^2 + R^2\,\sinh^2\left(t/R\right)\, d\phi^2.
\ee
Then, the de Sitter spacetime \eqref{metricdS2} has nonvanishing curvature components 
\bse\label{curvature}
\bea
R^t_{\phi t \phi} & = &  - R^t_{\phi \phi t} =  \sinh^2 (t/R),\\
R^\phi_{t \phi t}  & = &  - R^\phi_{t t \phi}  =  -1/R^2.
\eea
\ese
For the rest of the paper, we set $R=1$ for simplicity.

\begin{figure}[htb]
\includegraphics[trim=150 80 130 90, clip]{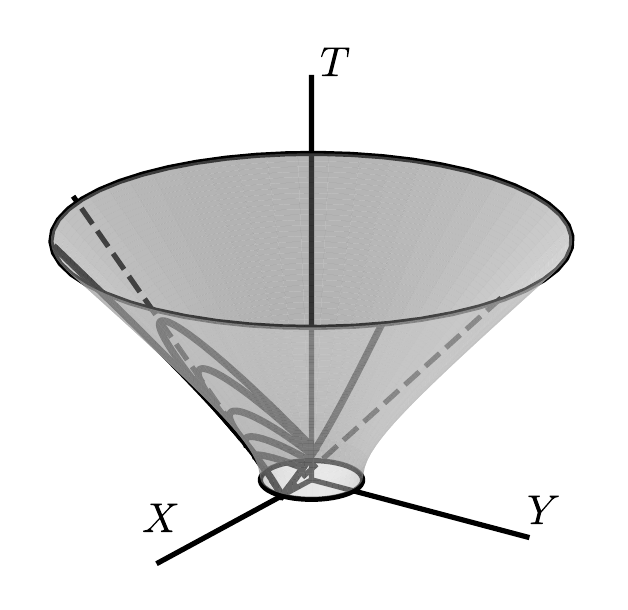}
\setlength{\abovecaptionskip}{5pt}
\setlength{\belowcaptionskip}{-10pt}
\caption{\label{fig:hyp}  The  portion  $0 \le T \le 5$  of the de Sitter hyperboloid (with $R=1$). The coordinate system $\{t,\phi\}$ covers the ``triangular wedge"  $X \ge 1$ bounded by the dashed lines.  The fiducial geodesic ($\phi = -\pi/3$) and the secondary geodesic ($\phi = \pi/3$) are the 
solid lines extending from $X=1$, $Y = T = 0$. Spacelike geodesics (solid curves) are superimposed on the hyperboloid  for  values of $t_0$ equal to $0.44$, $0.56$, $0.64$, $0.69$,  and $0.71$. The spacelike geodesics connect the fiducial and secondary geodesics. 
}
\end{figure}

As shown in Fig.~\ref{fig:hyp}, 
the coordinates $\left\{t, \phi\right\}$ cover the  ``triangular wedge" $X\ge 1$ of the hyperboloid (\ref{hyperboloid}).  
A curve of constant $t$ 
is the intersection of the hyperboloid with a plane $X = {\rm const} > 1$.  
A curve of constant $\phi$ is the intersection of the hyperboloid with a plane $T/Y = {\rm const}$ with $-1 < T/Y < 1$. 
The constant $\phi$ curves are timelike geodesics. 
Note that, in spite of its name, $\phi$ is not an angular coordinate. 
Rather, $\phi$ takes its values on the real number line. 
In Fig.~\ref{fig:hyp} the edges of the triangular wedge coincide with $\phi = \pm\infty$. 
The edges (shown as dashed lines) lie at the intersection of the hyperboloid with the plane $X = 1$.  
 
\begin{figure}[htb]
\includegraphics[trim=150 80 130 90, clip]{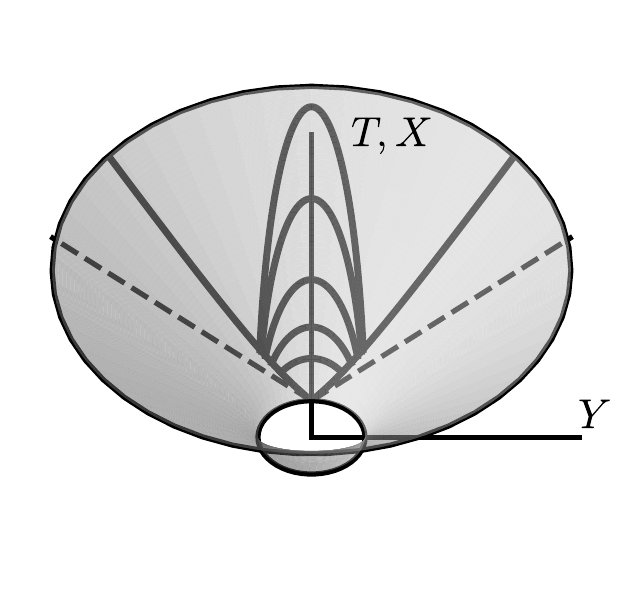}
\setlength{\abovecaptionskip}{5pt}
\setlength{\belowcaptionskip}{-10pt}
\caption{\label{fig:hypnewview}  The de Sitter hyperboloid of Fig.~\ref{fig:hyp}, with the ``camera"  placed in the $T$--$X$ plane at an angle of 45 degrees below the $X$--$Y$ plane.  From this view, the $T$ and $X$ axes appear to coincide. 
}
\end{figure}

 Figure \ref{fig:hypnewview} is 
identical to Fig.~\ref{fig:hyp}, but  viewed from a different perspective. In  both Figs.~\ref{fig:hyp} and \ref{fig:hypnewview} the dashed lines are the boundaries of the $\{t,\phi\}$ coordinate system. The solid curves are the timelike and spacelike geodesics discussed in the next section. 

\section{Timelike $\&$ Spacelike Geodesics}\label{sec4}

In the geometry of Eq.~\eqref{metricdS2}, the geodesic equations are 
\bse\label{geodesiceqdS2v1}
\bea
\frac{d^2 t}{d\lambda^2} + \sinh t\, \cosh  t\, \left(\frac{d\phi}{d\lambda}\right)^2 & = & 0,\,\\
\frac{d^2 \phi}{d\lambda^2} +2\, \coth t\, \frac{d t}{d\lambda} \,\frac{d \phi}{d\lambda}& = & 0,
\eea
\ese
where $\lambda$ is a general affine parameter.
We can see from inspection that curves of constant $\phi$ are timelike geodesics 
\bse\label{timelikegeoSOLdS2}
\bea
t(\tau) & = & \tau,\quad   \phi(\tau) = \phi_1, \,\, \left(\text{fiducial} \right)\\
t(\tau) & = & \tau, \quad \phi(\tau) = \phi_2,\,\, \left(\text{secondary} \right) \eea
\ese
where $\lambda = \tau$ is proper time. 
As illustrated in Figs.~\ref{fig:hyp} and \ref{fig:hypnewview}, these geodesics fan out from the crossing point defined by $T=Y=0$ and $X=1$, which is 
at the ``throat" of the hyperboloid. 
At the crossing point, the geodesics 
pass each other  with nonzero relative speed. 

To compute the relative acceleration in the isochronous correspondence, we choose points $P$ and $\bar P$ on the fiducial and secondary geodesics that have common values of proper time. 
(That is, common values of coordinate 
$t$.)
We construct a spacelike geodesic from $P$ to $\bar P$ by setting
$\lambda = s$ in the geodesic equations \eqref{geodesiceqdS2v1}, where 
$s$ is an affine parameter that runs from $0 \le s \le 1$. 
The spacelike geodesic, which we denote 
\be\label{spacelikegeo}
\xi^\mu(s) = \left(t(s),\phi(s)\right) ,
\ee
connects the fidicual geodesic at $s=0$ to the secondary geodesic at $s=1$. 
Let $t_0$ denote the $t$--coordinate value at $P$ and $\bar P$. 
Then, the boundary conditions for the spacelike geodesic are
\bse\label{BCtemp}
\bea
 t(s=0) & = & t_0, \quad  \phi(s=0) = \phi_1,\\
 t(s=1) & = & t_0, \quad \phi(s=1) = \phi_2 .
\eea
\ese
For simplicity, we will set $\phi_1 = -\phi_0$ and $\phi_2 = \phi_0$. 
Since de Sitter spacetime is symmetric under rotations about the $T$--axis, the spacelike geodesic will be symmetric about $\phi = 0$. 
In particular, the function $t(s)$ should reach an extremum at the midpoint, $s=1/2$, where $\phi = 0$.
This allows us to replace the boundary conditions~(\ref{BCtemp}b) with
\be\label{dtdsathalf}
\frac{dt(s)}{ds}\Bigg|_{s=1/2} = 0 , \quad \phi(s=1/2) = 0 .
\ee
These boundary conditions are easier to implement than Eqs.~\eqref{BCtemp} because the solution $t(s)$  switches branches from $dt/ds > 0$ to $dt/ds < 0$ at $s=1/2$. 

The geodesic equations~\eqref{geodesiceqdS2v1} with $\lambda = s$ and boundary conditions~(\ref{BCtemp}a) 
and~\eqref{dtdsathalf} are satisfied by
\bse\label{Realsolutions}
\bea
t(s) & = & \ln \left[ x + \sqrt{x^2 -1}\,\, \right],\\
\phi(s) & = &  \frac{1}{2} \ln \left[\frac{z + 1}{z -1} \right],
\eea\ese
where 
\bse\bea
x & \equiv &  -\frac{\sqrt{\ell^2 +  k^2}}{ k}\, \sin \left[ k(b-s) \right],\\
z & \equiv & \frac{\ell}{ k}\, \tan\left[ k(b-s) \right] .
\eea
\ese
The constant $k$ is defined by 
\be\label{kvalue}
k   = \arccos\,\biggl\{ 1 - 2\sinh^2(\phi_0) \sinh^2(t_0) \biggr\} ,
\ee
and the remaining constants are 
\bse\label{constantsofint}
\bea
b & = & \frac{1}{2}\left[1{ -} \frac{\pi}{ k}\right], \\
\ell  & = &  k\, \coth\left(\phi_0 \right)\, \tan\left(\frac{k}{2}\right) .
\eea\ese
Together, Eqs.~\eqref{Realsolutions} through (\ref{constantsofint}) describe the first half of the spacelike geodesic (the half with $0 \le s \le 1/2$) 
that spans the fiducial and secondary geodesics \eqref{timelikegeoSOLdS2} in the isochronous correspondence.

The constant $\ell$ in the spacelike geodesic is the constant of motion associated with the rotational symmetry of Eq.~\eqref{metricdS2}, with 
Killing vector field $\partial/\partial\phi$. 
The constant $k$ determines the norm of the tangent vector $d \xi^\mu/ds$ along the spacelike geodesics:
\be\label{constant1}
k^2  =  g_{\alpha \beta}\, \frac{d \xi^\alpha(s)}{ds}\, \frac{d \xi^\beta(s)}{ds}.  
\ee
If we had chosen $k=1$, as is customary for a spacelike geodesic, then the tangent vectors $d\xi^\kappa/ds$ would always have unit length. 
The appearance of $k$ in Eq.~\eqref{constant1} allows us to fix 
the parameter distance such that \mbox{$0 \le s \le 1$}. 
This ensures that the separation vector \eqref{Skappa} is proportional to the proper distance between the fiducial and secondary geodesics. 

Figures~\ref{fig:hyp} and \ref{fig:hypnewview} show the spacelike geodesics \eqref{Realsolutions} connecting the timelike geodesics~(\ref{timelikegeoSOLdS2}a,b) at the same value of affine parameter, for several distinct values of $t_0 = \tau$. 
These curves are superimposed on the triangular wedge of the de Sitter hyperboloid, which is the region covered by the coordinates $\left\{t, \phi\right\}$ of Eq.~\eqref{metricdS2}. 
In this picture the fiducial and secondary geodesics are defined by $\phi = -\pi/3$ and $\phi = \pi/3$, 
respectively. 
Thus, the connecting spacelike geodesics have $\phi_0 = \pi/3$. 

\begin{figure}[htb]
\includegraphics[trim=150 80 130 90, clip]{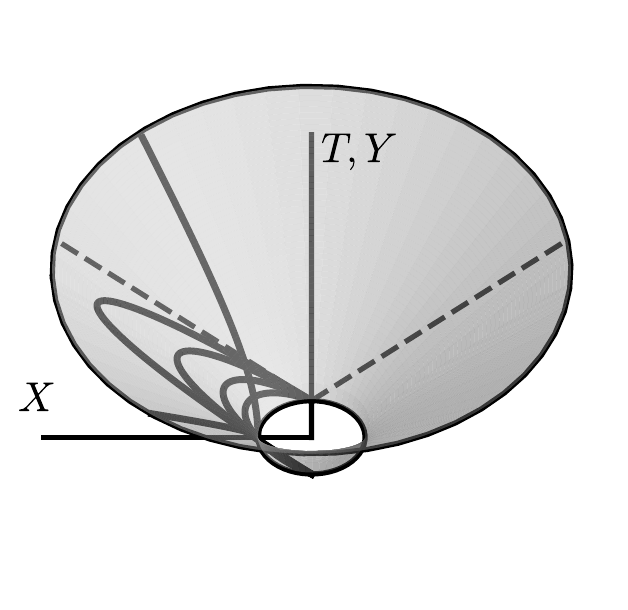}
\setlength{\abovecaptionskip}{5pt}
\setlength{\belowcaptionskip}{-10pt}
\caption{\label{fig:hypagain}   The spacelike geodesics (solid curves) are related by Lorentz boosts in the $T$--$X$ plane. These geodesics remain 
outside the light cone (dashed lines) emanating from the event  $X = T = 0$, $Y=1$.  The secondary timelike geodesic (solid line) 
crosses the light cone at the event $\{t_0^{\mathrm max}, \phi_0\}$. For $|t_0| > t_0^{\mathrm max}$, the fiducial and secondary geodesics 
cannot be connected by a spacelike geodesic. 
}
\end{figure}

The spacelike geodesics can be obtained by applying a Lorentz boost $X\partial/\partial T + T \partial/\partial X$ to the ``throat" $T=0$ of the hyperboloid, as shown in Fig.~\ref{fig:hypagain}. 
This boost leaves the points $Y=\pm 1$ unchanged. 
By considering all boosts of the ``throat", we obtain a sequence of spacelike geodesics that populate the region of spacetime that is 
exterior to the future and past light cones emanating from  $X=T=0$, $Y=\pm 1$. 
(The ``elsewhere" region of these light cones.) 
The boundaries of this region are the null 
curves defined by $T = \lambda$, $X = \lambda$, $ Y=\pm 1$. 

In Fig.~\ref{fig:hypagain}, the spacelike geodesics are extended beyond the fiducial and secondary geodesics until they reach the throat ($T=0$) of the hyperboloid. The ``camera" is placed in the $T$--$Y$ plane (so the $T$ and $Y$ axes appear to coincide) at an angle of $45$ degrees below the $X$--$Y$ plane. 

The argument above shows that the spacelike geodesics that are symmetric about the $T$--$X$ plane never cross the null curves 
$T = \lambda$, $X = \lambda$, $ Y=\pm 1$.
If the timelike geodesics~\eqref{timelikegeoSOLdS2} cross these null curves, then they cannot be connected by a spacelike geodesic. 
This will occur for large values of $t_0 = \tau$, values greater than some maximum $t_0^{\mathrm{max}}$. 

We can compute $t_0^{\mathrm{max}}$ by finding the point where the secondary geodesic $\phi = \phi_0$ crosses the null curve $T = \lambda$, $X = \lambda$, $ Y= 1$.
That is, we set 
\bse\bea
    T & = & \sinh(t_0) \cosh(\phi_0) = \lambda, \\
    X & = & \cosh(t_0) = \lambda, \\
    Y & = & \sinh(t_0) \sinh(\phi_0) = 1,
\eea\ese
and solve for $t_0$. 
This yields
\be\label{t0maxeqn}
    t_0^{\mathrm{max}} = \arcsinh\,\left[1/\sinh(\phi_0)\right].
\ee
The fiducial and secondary timelike geodesics cannot be connected by a spacelike geodesic (in the isochronous correspondence) unless 
$|t_0| < t_0^{\mathrm{max}}$.

The limit on $t_0$ can also be found from the 
solution~\eqref{Realsolutions} for the spacelike geodesics. 
Equation~\eqref{kvalue} for $k$ implies  that $\cos(k) = 1$ at $t_0=0$, and $\cos(k)$ decreases as $t_0$ increases. 
When  $\cos(k)$ drops below $-1$, $k$ is no longer real. This occurs 
when 
\be
    1 - 2\sinh^2(\phi_0)\sinh^2(t_0) = -1,
\ee
which has the solution given in Eq.~\eqref{t0maxeqn}. 
For the value $\phi_0=\pi/3$ used in Figs.~\ref{fig:hyp}, \ref{fig:hypnewview} and \ref{fig:hypagain}, we have $t_0^{\mathrm{max}} \simeq 0.733$.

\vspace{0.5cm}
\section{Geodesic Deviation in de Sitter Spacetime}\label{sec5}

Using the particular solutions for the spacelike geodesics found in Sec.~\ref{sec4}, the separation vector from  Eq.~\eqref{Skappa} has components
\bea\label{SdS2}
S^\kappa(\tau)  & \equiv & \frac{d \xi^\kappa(s)}{ds}\bigg|_{s=0} = \left(t^{\prime}(s), \phi^{\prime}(s) \right)\big|_{s=0},\nono\\
& = & \left(\sqrt{ \frac{\ell^2(\tau)}{\sinh^2 \tau} -k^2(\tau) },\, \frac{\ell(\tau)}{\sinh^2 \tau}\right),
\eea
where a prime denotes a derivative with respect to $s$.
Note that on the right--hand side, we have
replaced $t_0$ with $\tau$ since $\tau = t_0$ along the fiducial geodesic.
We have also suppressed all dependence on $\phi_0$ for simplicity. 
Likewise, the components of the tangent vector to the 
fiducial geodesic are
\be\label{TdS2}
T^\nu  {\equiv}   \left(\dot{t}(\tau) ,\dot{\phi}(\tau)  \right) = (1,0),
\ee
where an overdot denotes a derivative with respect to $\tau$.

Armed with Eqs.~\eqref{SdS2} and~\eqref{TdS2}, we can now compute the components of the relative velocity $DS^\kappa/d\tau$ between the fiducial and secondary geodesics. 
In de Sitter spacetime, these components are given by
\bse\label{velocity}
\bea \nono
&&\frac{DS^t(\tau)}{d\tau}  = \left(\ell(\tau )^2\,
   \text{csch}^2(\tau )-k(\tau )^2\right)^{-1/2} \times\\ 
&&  
\biggl\{\ell(\tau )\, \text{csch}^2(\tau)\, \left[\dot{\ell}(\tau )-\ell(\tau )\,
   \coth (\tau )\right]
   -k(\tau )\, \dot{k}(\tau )\biggr\}, \qquad\quad \\
&& \frac{DS^\phi(\tau)}{d\tau} = 
\text{csch}^2(\tau )\, \left[\dot{\ell}(\tau )-\ell(\tau)\, \coth (\tau )\right],
\eea
\ese
Likewise, we have 
\begin{widetext}
\bse\label{acceleration}
\bea\notag
A^t(\tau) & = & 
    \frac{1}{4 \,\left(\ell(\tau )^2\,
   \text{csch}^2(\tau )-k(\tau )^2\right)^{3/2}} \biggl\{4 \,\left(k(\tau )^2-\ell(\tau )^2\, \text{csch}^2(\tau
   )\right)\times\\\nono 
   && \quad \quad \left[k(\tau ) \,\ddot{k}(\tau )+\dot{k}(\tau
   )^2+\text{csch}^2(\tau ) \left(-\left(\dot{\ell}(\tau )^2+\ell(\tau )
   \left(\ddot{\ell}(\tau )-4\, \coth (\tau )\, \dot{\ell}(\tau )\right)+\ell(\tau
   )^2 \left(3\, \text{csch}^2(\tau
   )+2\right)\right)\right)\right]\\
   && \quad \quad \quad \quad \quad \quad \quad \quad \quad \quad \quad \quad \quad \quad \quad \quad  -4 \left(k(\tau ) \,\dot{k}(\tau
   )+\ell(\tau )\, \text{csch}^2(\tau ) \left(\ell(\tau )\, \coth (\tau
   )-\dot{\ell}(\tau )\right)\right)^2\biggr\},\\
A^\phi(\tau) & = & 
\text{csch}^2(\tau )\, \left[\ddot{\ell}(\tau )-2\, \coth (\tau )\,\dot{\ell}(\tau
   )+\ell(\tau )+2\,\ell(\tau )\, \text{csch}^2(\tau )\right],
\eea
\ese
\end{widetext}
for the components of the relative acceleration \mbox{$A^\kappa \equiv D^2S^\kappa/d\tau^2$} between the fiducial and secondary geodesics.
In Eqs.~\eqref{velocity} through~\eqref{acceleration}, $k(\tau)$ and $\ell(\tau)$ are given by Eqs.~(\ref{kvalue}) and~(\ref{constantsofint}b), respectively, with $t_0$ replaced by $\tau$.

Now we compute the right--hand side of the generalized geodesic deviation equation \eqref{ggdedS2v1}. The $t$--component includes the terms 
\begin{widetext}
\bse\label{ggdetcompdS2}
\bea
R^t_{\lambda \mu \nu}\,T^\lambda\,S^\mu\,T^\nu & = &  0,\\
2\, R^t_{\lambda \mu \nu}\, \frac{DS^\lambda}{d\tau}\, S^\mu\, T^\nu & = & 2\, \sinh^2(\tau)\,\frac{DS^\phi(\tau)}{d\tau}\,\left [-S^\phi(\tau)\right],\\
\frac{2}{3} R^t_{\lambda \mu \nu}\,\frac{DS^\lambda}{d\tau}\, S^\mu\, \frac{DS^\nu}{d\tau} & = &  \frac{2}{3}\, \sinh^2(\tau)\, \frac{DS^\phi(\tau)}{d\tau}\, \left[S^t\, \frac{DS^\phi(\tau)}{d\tau} - S^\phi(\tau)\, \frac{DS^t(\tau)}{d\tau} \right].
\eea
\ese
Similarly, the $\phi$--component of Eq.~\eqref{ggdedS2v1} includes 
\bse\label{ggdephicompdS2}
\bea
R^\phi_{\lambda \mu \nu}\,T^\lambda\,S^\mu\,T^\nu & = &  - S^\phi(\tau),\\
2\, R^\phi_{\lambda \mu \nu}\, \frac{DS^\lambda}{d\tau}\, S^\mu\, T^\nu & = & - 2\, \frac{DS^t(\tau)}{d\tau}\, S^\phi(\tau),\\
\frac{2}{3} R^\phi_{\lambda \mu \nu}\,\frac{DS^\lambda}{d\tau}\, S^\mu\, \frac{DS^\nu}{d\tau} & = & -\frac{2}{3}\, \frac{DS^t(\tau)}{d\tau}\, \left[S^\phi(\tau)\,\frac{DS^t(\tau)}{d\tau} - S^t(\tau)\, \frac{DS^\phi(\tau)}{d\tau} \right].
\eea
\ese
\end{widetext}
We have carried out the implied sums over repeated indices. 

From Eqs.~(\ref{velocity}) through~(\ref{ggdephicompdS2}), we want to construct manifestly covariant scalars that are real and positive. 
Obtaining such quantities will allow us to make definitive statements about the GDE and the GGDE in de Sitter spacetime that all observers will agree upon.  
First, define the following vector fields
\bse\label{vectorfields}
\bea
V^\kappa & \equiv & - R^\kappa_{\lambda \mu \nu} T^\lambda S^\mu T^\nu,\\
W^\kappa & \equiv &  - R^\kappa_{\lambda \mu \nu}  \,\biggl[
T^\lambda S^\mu T^\nu + 2\,\frac{DS^\lambda}{d\tau}\, S^\mu\, T^\nu \nono\\
& & \qquad\qquad +  \frac{2}{3}\, \frac{D S^\lambda}{d\tau}\, S^\mu\,\frac{DS^\nu}{d\tau}\biggr].
\eea
\ese
Then, the geodesic deviation equation~\eqref{GDE} and the generalized geodesic deviation equation~\eqref{ggdedS2v1} can be recast as
\bse\label{GDEGGDE}
\bea
A^\kappa & = &  V^\kappa + \mathcal{O}\left(S, {DS}/{d\tau}\right)^2,\\
A^\kappa  & = & W^\kappa + \mathcal{O}(S)^2 ,
\eea
\ese
respectively, where $A^\kappa \equiv D^2S^\kappa/d\tau^2$.

Let $|A| \equiv \sqrt{|g_{\kappa\gamma} A^\kappa A^\gamma|}$ denote the magnitude of the vector $A^\kappa$; similarly, let $|V|$, $|W|$, and $|S|$ denote the magnitudes of $V^\kappa$, $W^\kappa$, and $S^\kappa$, respectively. 
Then the geodesic deviation equation~(\ref{GDEGGDE}a) implies 
\be\label{gdeAoverS}
    \frac{|A|}{|S|} = \frac{|V|}{|S|} + \mathcal{O}\left(S,\frac{DS}{d\tau}\right)^1, 
\ee
and the generalized geodesic deviation equation~(\ref{GDEGGDE}b) gives
\be\label{ggdeAoverS}
     \frac{|A|}{|S|} = \frac{|W|}{|S|}  + \mathcal{O}(S)^1.
\ee
Note that the difference $|A|/|S| - |V|/|S|$ depends on relative velocity terms $DS^\kappa/d\tau$, but the difference $|A|/|S| - |W|/|S|$ does not.

More precisely, we can compare the expressions for $V^\kappa$ and $W^\kappa$ to see that the error terms in the geodesic deviation equation~(\ref{GDEGGDE}a) include  
\be\label{GDEerrorterms}
     \Delta^\kappa = -  R^\kappa_{\lambda \mu \nu}  \biggl[ 2\frac{DS^\lambda}{d\tau} S^\mu
 T^\nu +  \frac{2}{3}\frac{DS^\lambda}{d\tau} S^\mu \frac{DS^\nu}{d\tau}\biggr]  ,
\ee
where $\Delta^\kappa \equiv W^\kappa - V^\kappa$. 
By computing the magnitude of $A^\kappa$ from Eqs.~\eqref{GDEGGDE}, we find that the $\mathcal{O}(S,DS/d\tau)^1$ errors in Eq.~\eqref{gdeAoverS} include
\be\label{errortermsingde}
    \frac{V_\kappa \,\Delta^\kappa}{|V|\,|S|} = - \frac{V_\kappa}{|V|} R^\kappa_{\lambda \mu \nu}  \biggl[ 2\frac{DS^\lambda}{d\tau} \frac{S^\mu}{|S|}
 T^\nu +  \frac{2}{3}\frac{DS^\lambda}{d\tau} \frac{S^\mu}{|S|} \frac{DS^\nu}{d\tau}\biggr]  .
\ee
These terms do not vanish in the limit as the separation $|S|$ goes to zero unless the rate of separation $DS^\kappa/d\tau$ is also vanishingly small.

To isolate the effects that the relative velocity terms have on the relative acceleration, we must 
consider timelike geodesics that are close to each other but moving at high relative velocity. 
In de Sitter spacetime, this occurs for the geodesics \eqref{timelikegeoSOLdS2} when they are near the crossing point 
at the throat of the hyperboloid. 
That is, as $\tau ( = t_0)$ tends to zero (see Fig.~\ref{fig:hyp}).

\begin{figure}[h!]
\centering
\includegraphics[width=80mm,scale=20.5]{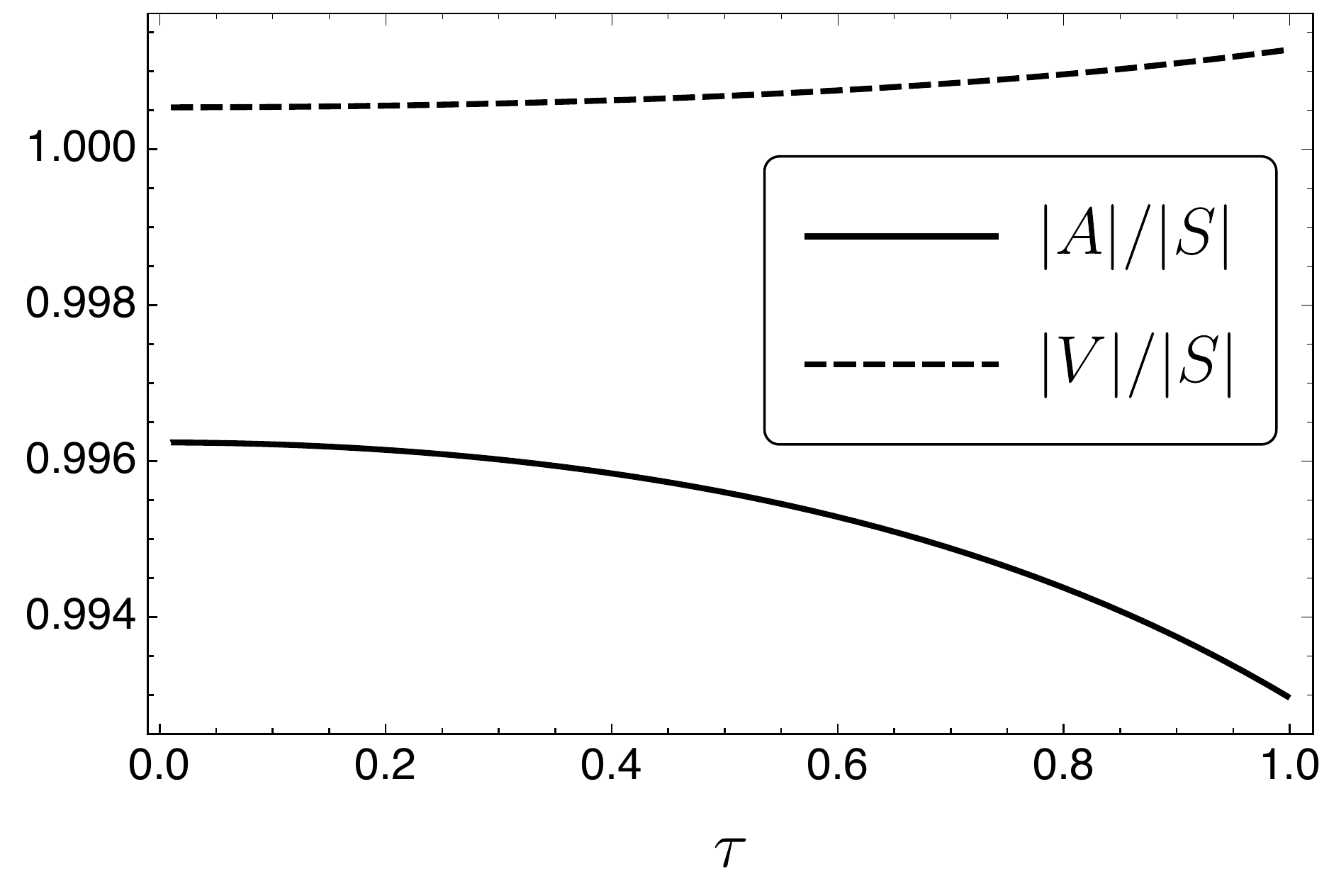}
\setlength{\abovecaptionskip}{5pt}
\setlength{\belowcaptionskip}{-10pt}
\caption{The geodesic deviation equation. The scalar $|V|/|S|$ from the GDE is compared to the scalar $|A|/|S|$ for $\phi_0 = \pi /96$.}
\label{fig:GDE}
\end{figure}

\begin{figure}[h!]
\centering
\includegraphics[width=80mm,scale=20.5]{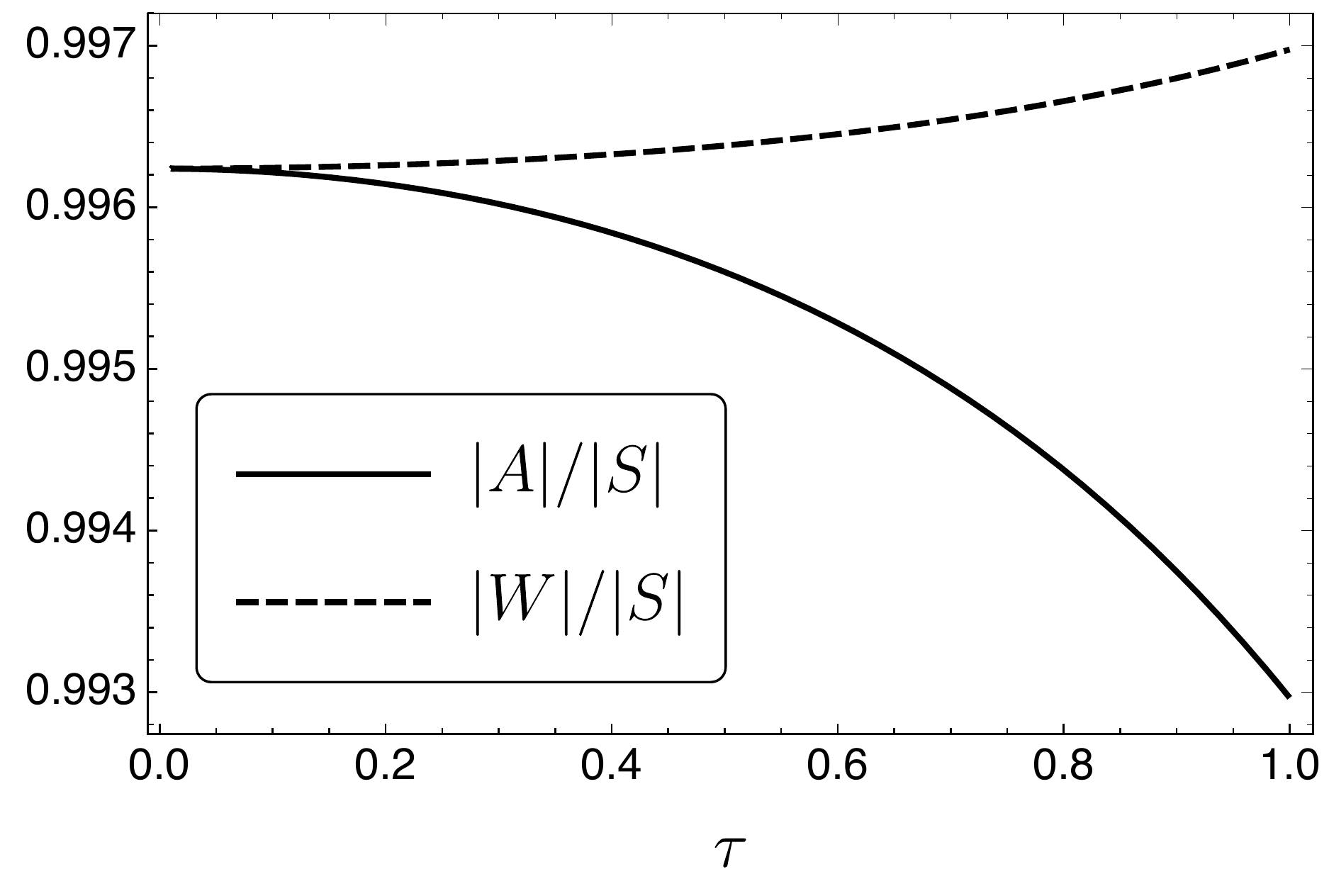}
\setlength{\abovecaptionskip}{5pt}
\setlength{\belowcaptionskip}{-10pt}
\caption{The generalized geodesic deviation equation. The scalar $|W|/|S|$ from the GGDE is compared to the scalar $|A|/|S|$ for $\phi_0 = \pi /96$.}
\label{fig:GGDE}
\end{figure}

Using the results from Eqs.~(\ref{velocity}) through~(\ref{ggdephicompdS2}), we plot the scalars $|A|/|S|$, $|V|/|S|$, and $|W|/|S|$ in Figures~\ref{fig:GDE} and~\ref{fig:GGDE}. 
In these figures the coordinate separation between the timelike geodesics is $\phi_2 = -\phi_1 = \phi_0 =\pi/96$. 

Figures~\ref{fig:GDE} and~\ref{fig:GGDE} illustrate the distinction between the GDE and the GGDE.
In Fig.~\ref{fig:GDE}, the scalar $|V|/|S|$ disagrees with the relative acceleration $|A|/|S|$ between timelike geodesics that are close ($\tau \to 0$) but moving with nonzero relative velocity.
Figure~\ref{fig:GGDE} shows that the scalar $|W|/|S|$ agrees with the relative acceleration as $\tau$ approaches zero. 

For the example shown in Figs.~\ref{fig:GDE} and~\ref{fig:GGDE}, with $\phi_0 =\pi/96$,  we can explicitly compute the effect of the terms~\eqref{errortermsingde}.  
First introduce a basis of orthonormal covectors,  $e_\mu^{\hat t} = (1,0)$ and $e_\mu^{\hat\phi} = (0,\sinh t)$. 
In this basis the components of the separation vector are 
\be
    S^{\hat\kappa}  = e^{\hat\kappa}_\kappa S^\kappa 
    \approx (0.00214 , 0.0655 ) \, \tau, 
\ee
in the limit of small $\tau$. 
Likewise, the orthonormal basis components of the relative velocity vector are
\be\label{relativevelocityvector}
    \frac{DS^{\hat\kappa}}{d\tau} \approx (0.00214, 0.0655) ,
\ee
and the components of the tangent vector are $T^{\hat\mu} = (1,0)$. 
The Riemann tensor has components $\pm 1$. 

Since the component $T^{\hat \phi}$ vanishes, the definition~(\ref{vectorfields}a) and the symmetries of Riemann tell us that the component $V^{\hat t}$ must be zero. 
It follows that $V^{\hat\kappa}/|V| = (0,1)$, and the errors in Eq.~\eqref{errortermsingde} come entirely from the $\hat\phi$ component of $\Delta^{\hat\kappa}/|S|$. 
In turn, from Eq.~\eqref{GDEerrorterms} we see 
that 
\be\label{Deltaphihat}
    \frac{\Delta^{\hat\phi}}{|S|} \approx \pm \left[ 2 \frac{DS^{\hat t}}{{d\tau}} \frac{S^{\hat\phi}}{|S|} 
    + \frac{2}{3} \left( \frac{DS^{\hat t}}{{d\tau}} \right)^2 \frac{S^{\hat\phi}}{|S|}  \right],
\ee
where the factor $\pm 1$ comes from a component of the Riemann tensor. 

Using the numbers from our example (with \mbox{$\phi_0 = \pi/96$}), we have \mbox{$S^{\hat\phi}/|S| \approx 1.0$} and \mbox{$DS^{\hat t}/d\tau \approx 0.00214$}. Putting these together in Eq.~\eqref{Deltaphihat} above  yields $\Delta^{\hat\phi}/|S| \approx \pm 0.0043$. 
This result agrees, in the limit as $\tau\to 0$, with the discrepancy seen in Fig.~\ref{fig:GDE} between the relative acceleration $|A|/|S|$ and the result $|V|/|S|$ that follows from the geodesic deviation equation.

This example shows explicitly that the GDE does not hold in the limit as the separation  goes to zero because the rate of separation remains finite. 
To be precise, in de Sitter spacetime~\eqref{metricdS2} with $R=1$, the Ricci scalar is $2$ and the scale of curvature is $\mathcal{R} \approx 1/\sqrt{2} \approx 0.71$.\footnote{${\mathcal R}$, like $|S|$, has dimensions of length. Units are suppressed in these calculations.}
The ratio $|S|/{\mathcal R} \approx 0.093\,\tau$ goes to zero in the limit as $\tau\to 0$. 
On the other hand, the rate of separation \eqref{relativevelocityvector} is not vanishingly small compared to the speed of light $c$ (which equals $1$).

Of course, rapidly diverging geodesics will only remain closely separated on timescales less than $\mathcal{R}/c$, implying that the GGDE approximation $A^\kappa \approx W^\kappa$ will eventually break down. 
We can see this in Fig.~\ref{fig:GGDE}, where the curves diverge as $\tau$ increases. 
As the timelike geodesics diverge away from the crossing point, 
the dashed curve becomes worse at predicting the relative acceleration because the separation between geodesics no longer satisfies \mbox{$|S|/\mathcal{R} \ll 1$}. 
In this regime the $\mathcal{O}(S)^2$ errors in~(\ref{GDEGGDE}b) are non-negligible, implying that the GGDE approximation $A^\kappa \approx W^\kappa$ no longer applies.

\section{Summary}\label{sec6}

We have articulated the difference between the GDE and the GGDE by explicitly computing the relative acceleration between timelike geodesics in two--dimensional de Sitter spacetime. 
This spacetime provides a rare and interesting example of a fully analytical calculation of the relative acceleration in the isochronous correspondence. 
The maximal symmetry of de Sitter spacetime provides constants of motion that allow us to explicitly solve the boundary value problem for the spacelike geodesic that connects two timelike geodesics at the same value of affine parameter.

As we approach the throat of the hyperboloid in Fig.~\ref{fig:hyp}, the separation between geodesics is small compared to $\mathcal{R}$ but their rates of separation are not small compared to $c$. 
The geodesics have crossed one another at high relative velocity. 
In this situation, the relative acceleration is correctly described by the GGDE~(\ref{GDEGGDE}b).  
The GDE~(\ref{GDEGGDE}a), which dominates the discussion of geodesic deviation in textbooks on general relativity, does not apply.

The methods of computing geodesic deviation used by
Refs.~\cite{Vines2015, Flanagan2019, Flanagan2020, Puetzfeld2016, Obukhov2019} require knowledge of fundamental bitensors.
These bitensors are known in a modicum of spacetimes, including the class of plane wave spacetimes~\cite{Harte2012} considered by Refs.~\cite{Flanagan2019, Flanagan2020}.
There may be other spacetimes with enough symmetries to permit analytical treatments of geodesic deviation.
But in most spacetimes, these types of calculations must be done numerically. 
Our work has the potential to inform future numerical investigations since it adds to the literature on the analytical calculations of geodesic deviation at higher order.

\begin{acknowledgements}
I.R.W. and J.D.B. acknowledge the support of the Hamilton Award from the Department of Physics and Astronomy at UNC Chapel Hill.
\end{acknowledgements}

\bibliography{bibv1}

\end{document}